\documentclass[11pt]{article}
	
	\usepackage{hhline}
	\newcommand{\blind}{0}
	\usepackage{amsmath, amssymb, amscd, amsthm, amsfonts}
    \usepackage{graphicx}
    \usepackage{hyperref}
    \usepackage{multirow}
    \usepackage{color, colortbl}
    \usepackage{amsfonts} 
    \usepackage{amssymb}
    \usepackage{breqn}
    \usepackage{adjustbox}
    \usepackage{alphabeta}
    
    \newcommand{\thickhline}{%
    \noalign {\ifnum 0=`}\fi \hrule height 1pt
    \futurelet \reserved }
    \usepackage{algorithmic}
    \usepackage{algorithm}
    \usepackage{alphabeta}
    
    \usepackage[symbol]{footmisc}

    \usepackage{fancyhdr}
    \fancypagestyle{mystyle}{%
    \fancyhead[LE,RO]{This work has been accepted for publication at the IISE Transactions}\fancyfoot{}
}%

    \usepackage{algorithmic}
    \usepackage[dvipsnames]{xcolor}
    
    \usepackage{array}
    \usepackage{makecell}
    \usepackage{mathtools}
    \makeatletter
    \renewcommand\section{\@startsection {section}{1}{\z@}%
                                       {-3.5ex \@plus -1ex \@minus -.2ex}%
                                       {2.3ex \@plus.2ex}%
                                       {\normalfont\fontfamily{phv}\fontsize{16}{19}\bfseries}}
    \renewcommand\subsection{\@startsection{subsection}{2}{\z@}%
                                         {-3.25ex\@plus -1ex \@minus -.2ex}%
                                         {1.5ex \@plus .2ex}%
                                         {\normalfont\fontfamily{phv}\fontsize{14}{17}\bfseries}}
    \renewcommand\subsubsection{\@startsection{subsubsection}{3}{\z@}%
                                        {-3.25ex\@plus -1ex \@minus -.2ex}%
                                         {1.5ex \@plus .2ex}%
                                         {\normalfont\normalsize\fontfamily{phv}\fontsize{14}{17}\selectfont}}
    \makeatother
	
	\usepackage{amsmath}
	\usepackage{graphicx}
	\usepackage{enumerate}
	\usepackage{natbib} 
	\usepackage{url} 
	
\begin{document}
	
	\newcolumntype{C}[1]{>{\centering\arraybackslash}p{#1}}
		
		\def\spacingset#1{\renewcommand{\baselinestretch}%
			{#1}\small\normalsize} \spacingset{1}
		
		\if0\blind
		{
			\title{\bf 
			STOCHOS: Stochastic Opportunistic Maintenance Scheduling For Offshore Wind Farms }
			\author{ Petros Papadopoulos, David W. Coit , Ahmed Aziz Ezzat \\
			Department of Industrial \& Systems Engineering, Rutgers University 
			}
			\date{}
			\maketitle
		} \fi
		
		\if1\blind
		{

            \title{\bf \emph{IISE Transactions} \LaTeX \ Template}
			\author{Author information is purposely removed for double-blind review}
			
\bigskip
			\bigskip
			\bigskip
			\begin{center}
				{\LARGE\bf \emph{IISE Transactions} \LaTeX \ Template}
			\end{center}
			\medskip
		} \fi
	\begin{abstract}
Despite the promising outlook, the numerous economic and environmental benefits of offshore wind energy are still compromised by its high operations and maintenance (O\&M) expenditures. On one hand, offshore-specific challenges such as site remoteness, harsh weather, transportation requirements, and production losses, significantly inflate the O\&M costs relative to land-based wind farms. On the other hand, the uncertainties in weather conditions, asset degradation, and electricity prices largely constrain the farm operator's ability to identify the time windows at which maintenance is possible, let alone optimal. 
In response, we propose STOCHOS, 
short for the \underline{stoc}hastic \underline{h}olistic \underline{o}pportunistic \underline{s}cheduler\textemdash a maintenance scheduling approach tailored to address the unique challenges and uncertainties in offshore wind farms. Given probabilistic forecasts of key environmental and operational parameters, STOCHOS optimally schedules the offshore maintenance tasks by harnessing the opportunities that arise due to favorable weather conditions, on-site maintenance resources, and maximal operating revenues. STOCHOS is formulated as a two-stage stochastic mixed integer linear program, which we solve using a scenario-based rolling horizon algorithm that aligns with the industrial practice. 
Tested on real-world data from the U.S. North Atlantic where several offshore wind farms are in-development,
STOCHOS demonstrates considerable improvements relative to prevalent maintenance benchmarks, across various O\&M metrics, including total cost, downtime, resource utilization, and maintenance interruptions. 
	\end{abstract}
			

	\spacingset{1.5} 
\thispagestyle{mystyle}

\section{Introduction}
\label{s:intro}
Offshore wind (OSW) is set to play a pivotal role in the global climate change mitigation efforts. As a clean source of energy, it has a significantly smaller environmental footprint than fossil fuels, while it enables access to stronger and steadier winds than those in-land, resulting in a higher and more reliable generation of electricity. Several countries have set multi-year targets for the adoption or the expansion of their OSW portfolios, with recent reports projecting a growth in its global capacity of at least $15$ times by $2040$ \citep{iea2019outlook_OW}. 

Despite the promising outlook, OSW still faces substantial technical challenges pertaining to the high expenditures associated with operating and maintaining a fleet of OSW turbines. Aside from the high upfront capital costs\textemdash which are largely due to the foundation construction and underwater cabling\textemdash a substantial portion of the total life cycle cost of OSW farms is attributed to operations and maintenance (O\&M) activities. Recent estimates suggest that O\&M activities performed throughout an OSW farm's lifetime contribute by more than 30\% to the total life cycle costs \citep{stehly20202018}.

\subsection{The need for opportunistic maintenance:} 
The high O\&M costs in OSW farms are driven by several unique challenges, including:

\hspace{-6.5mm}\textit{(C1)} \underline{OSW farm accessibility:} The ability of the maintenance crew to safely access the OSW farm is frequently prohibited by harsh metocean conditions (e.g., high wind speeds and wave heights) \citep{gilbert2021probabilistic}. Our analysis of the metocean conditions in the designated OSW areas in the NY/NJ Bight (where several OSW projects are currently in-development) suggests that an OSW turbine sited at that location can be inaccessible for $\sim$\hspace{-0.02mm}$56$\% of its operational time due to unsafe wind and/or wave conditions \citep{papadopoulosIEEE}. 

\hspace{-6.5mm}\textit{(C2)} \underline{Transportation and maintenance requirements:} OSW turbines are installed at remote locations that are accessed via repair vessels or helicopters, typically chartered from private O\&M contractors. Crew and equipment transport alone contributes $\sim$\hspace{0.2mm}$28$-$73\%$ of the OSW O\&M costs \citep{carroll2017availability,dalgic2015advanced}. Moreover, the harsh weather conditions at OSW farms further accelerate asset degradation and failure rates \citep{carroll2016failure}, which directly translate into frequent maintenance visits. Those high maintenance requirements largely inflate the maintenance setup costs relative to onshore wind farms.

\hspace{-6.5mm}\textit{(C3)} \underline{Revenue losses:} Modern OSW turbines have $\sim$\hspace{-0.1mm}$2$-$3$ times higher capacity than their land-based counterparts \citep{golparvar2021surrogate}. 
A shutdown of such an ultra-scale turbine results in large revenue losses, especially at times of strong winds or favorable market prices. 

Those offshore-specific challenges (\textit{C1}-\textit{C3}) make the conventional maintenance scheduling approaches\textemdash which may be well-suited for land-based operations\textemdash sub-optimal, motivating the need for an 
``\textit{opportunistic}'' approach for OSW maintenance scheduling. 
The essence of opportunistic maintenance is to incentivize the grouping of maintenance tasks (that would have been otherwise scheduled independently) at times of ``maintenance opportunity,'' in order to reduce the number of maintenance visits and aggregate them at times when it is most economical for the OSW farm operator \citep{ren2021offshore}.
In light of \textit{C1}-\textit{C3}, those opportunities can be one of the following: (\textit{i}) access-based opportunities: grouping maintenance tasks at times of OSW farm access to leverage the (often scattered) periods of provisional accessibility, hence minimizing maintenance delays, interruptions, and downtime; (\textit{ii}) resource-based opportunities: grouping maintenance tasks to leverage maintenance resources that are already on-site (e.g., vessels, equipment, crew) in order to share the setup cost among multiple turbines and maximize resource utilization; and (\textit{iii}) revenue-based opportunities: grouping maintenance tasks at times when production losses would be minimal (e.g., at times of anticipated low winds and/or low market prices).

However, our survey of the literature (see Table \ref{tab:references}) reveals that the overwhelming majority of research in wind farm opportunistic maintenance solely focuses on resource-based opportunities, which are often referred to in the literature as ``economic dependencies'' \citep{shafiee2015opportunistic, wang2019optimizing, ding2012opportunistic, sarker2016minimizing, ma2022multi, besnard2010approach, ko2017condition}. A small fraction of those efforts partially account for revenue losses by incentivizing the grouping of maintenance tasks at periods of low power production \citep{besnard2009optimization, yildirim2017integrated, song2018integrated}. A separate line of research focuses on forecasting access-based opportunities \citep{yang2020operations, lubing2019opportunistic, taylor2018probabilistic, zhang2021uncertain}, but largely overlooks those based on economic dependencies and/or revenue losses. 

Barring few recent efforts \citep{mazidi2017strategic,besnard2011stochastic,fallahi2022chance,papadopoulosIEEE}, there is a lack of approaches that holistically account for multiple components of maintenance opportunity. \cite{papadopoulosIEEE} show that such a holistic approach can lead to major O\&M cost savings due to the inter-dependencies between the three opportunistic components, making their combined impact on O\&M costs multiplicative rather than additive. For instance, both access- and revenue-based opportunities are weather-related, while resource-based opportunities are only possible when the OSW farm is accessible. However, a main limitation in the work of \cite{papadopoulosIEEE} is its inherent assumption that perfect knowledge about key environmental and operational parameters is available to the maintenance planner.  
\begin{table}[h]
    \caption{The contribution of STOCHOS to the OSW opportunistic maintenance literature} 
    \label{tab:references}
    \renewcommand{\arraystretch}{.52}
    \centering
    
    \begin{tabular}{| c | C{1cm} | C{1cm} | C{1cm} | C{1cm} | C{1cm} | C{1cm} | C{1cm} |}        
        \hline
        \multirow{2}{*}{ \textbf{Research Effort}} & \multicolumn{3}{c|}{ \textbf{\thead{Opportunities}  }}  & \multicolumn{4}{c|}{ \textbf{\thead{Uncertainties}}} \\\cline{2-8}
             
        &  \rotatebox{90}{\small Access ~~} & \rotatebox{90}{\small Revenue ~} &  \rotatebox{90}{\small Economic ~} &  \rotatebox{90}{\small Access ~~} & \rotatebox{90}{\small Production ~} &  \rotatebox{90}{\small Elec. prices ~~} &  \rotatebox{90}{\small Degradation~~}  \\
        \hline
      
        \cite{besnard2009optimization} & - & \checkmark & \checkmark & - & - & - & -  \\
       
        \cite{shafiee2015opportunistic} & - & - & \checkmark & - & - & - & \checkmark  \\
       
        \cite{wang2019optimizing} & - & - & \checkmark & - & - & - & \checkmark  \\
       
        \cite{ding2012opportunistic} & - & - & \checkmark & - & - & - & \checkmark  \\
       
        \cite{sarker2016minimizing} & - & - & \checkmark & - & - & - & \checkmark  \\
        
        \cite{ma2022multi} & - & - & \checkmark & - & - & - & \checkmark \\
        
        \cite{ko2017condition} & - & - & \checkmark & - & - & - & \checkmark  \\
       
        \cite{yildirim2017integrated} & - & \checkmark & \checkmark & - & \checkmark & \checkmark & \checkmark  \\
        
        \cite{mazidi2017strategic} & \checkmark & \checkmark & \checkmark & - & \checkmark & \checkmark & -  \\
       
        \cite{song2018integrated} & - & \checkmark & \checkmark & - & - & - & \checkmark  \\
        
        \cite{fallahi2022chance} & - & \checkmark & \checkmark & - & \checkmark & \checkmark & \checkmark  \\
        
        \cite{besnard2011stochastic} & \checkmark & \checkmark & \checkmark & \checkmark & \checkmark & - & -  \\
       
        \cite{besnard2010approach} & - & - & \checkmark & - & - & - & -  \\

        \cite{yang2020operations} & \checkmark & \checkmark & - & - & - & - & \checkmark  \\
       
        \cite{lubing2019opportunistic} & \checkmark & - & - & \checkmark & - & - & \checkmark  \\

        \cite{taylor2018probabilistic} & \checkmark & \checkmark & - & \checkmark & - & - & -  \\

        \cite{papadopoulosIEEE} & \checkmark & \checkmark & \checkmark & - & - & - & -  \\
       
        \textbf{STOCHOS} & \checkmark & \checkmark & \checkmark & \checkmark & \checkmark & \checkmark & \checkmark  \\
        
        
\hline
    \end{tabular}
\end{table}

\subsection{The value of recognizing uncertainty in OSW maintenance}
Several prior studies have advocated the importance of uncertainty modeling in wind farm maintenance planning \citep{byon2010optimal, perez2015multi}. We argue that, for an opportunistic maintenance approach, the need to recognize uncertainty is even more pressing because the imperfect knowledge about key environmental and operational parameters can largely compromise (or may even reverse) the economic gains of an opportunistic approach. 

The main reason is that those parameters, which are uncertain and difficult to predict, are the primary determinants of the ``opportunity windows'', i.e. the times at which maintenance grouping is economically desirable, including: (\textit{i}) uncertainty of metocean conditions (wind speed and wave height), which are the main determinants of OSW farm access (access- and resource-based opportunities), as well as the power production (revenue-based opportunities), (\textit{ii}) uncertainty of electricity prices, which govern the generated revenues (revenue-based opportunities), and (\textit{iii}) uncertainty in asset degradation, which affects the turbine's downtime and production losses (revenue- and resource-based opportunities). 

This suggests that an opportunistic maintenance approach, despite being desirable in essence, could have conflicting effects: On one hand, if (near) perfect knowledge of the uncertain parameters listed above is available, then the maintenance planner can precisely anticipate the windows of opportunity, and hence, fully harness the economic gains of an opportunistic approach. On the other hand, incorrect information about such parameters can have an adverse impact: Multiple maintenance tasks would be grouped at a time when maintenance may not even be feasible, let alone optimal, leading to significant maintenance delays and excessive downtime. In such situations, even an ``unopportunistic'' approach that individually schedules maintenance actions would have possibly been more cost-efficient. 
\begin{figure}[h]
    \centering
    \includegraphics[width=0.7\linewidth]
    {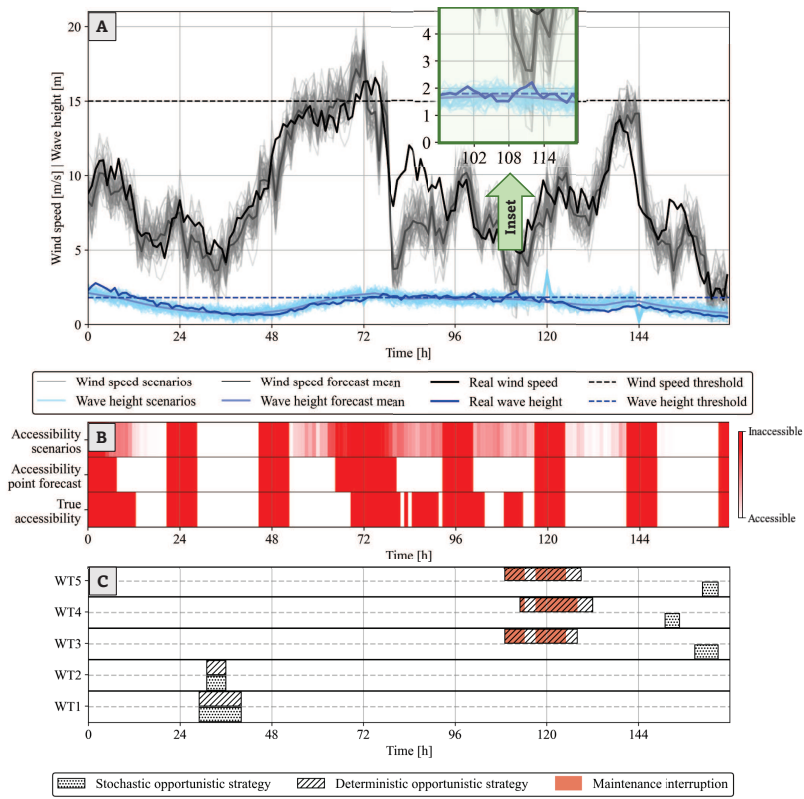}
    \caption{
    (A) Wind speed and wave height data, forecasts, along with $50$ scenarios. (B) Accessibility states. (C) Maintenance schedules of a deterministic and a stochastic strategy.}
    \label{fig:intro_uncertainty}
\end{figure}

Figure \ref{fig:intro_uncertainty} illustrates the value of considering uncertainty in OSW opportunistic maintenance with a realistic example, where the wind speed and wave height data, forecasts, and scenarios are displayed in Figure \ref{fig:intro_uncertainty}-A. 
Figure \ref{fig:intro_uncertainty}-B conveys information about the ``access state'';  the OSW farm is inaccessible when the wave height and/or the wind speed exceed their respective safety thresholds. Specifically, the top row of Figure \ref{fig:intro_uncertainty}-B shows the scenario-based probability of access, whereas the middle and bottom rows show the accessibility as a binary variable informed by the forecasts and real data, respectively. Figure \ref{fig:intro_uncertainty}-C is a Gantt chart, showing the maintenance schedules obtained by two optimization models: a deterministic (dotted bars) and a stochastic (hashed bars) model. The $x$-axis denotes the planning horizon (in hours) and the $y$-axis denotes the wind turbine (WT) index (here, we have $5$ WTs). Note that both optimization models are opportunistic, i.e., they both leverage access-, resource-, and revenue-based opportunities. Yet, the main difference is that the former only takes as input the point forecasts of the environmental and operational parameters listed above, while the latter makes use of the full probabilistic distribution of the forecasts to generate uncertainty scenarios as inputs to the stochastic optimization. 

In the case of the deterministic model which assigns complete confidence to the point forecasts, \textit{small forecasting errors can incur significant maintenance costs}. Consider the fifth day ($96$-$120$-h range) where the point forecasts under-predict the wave height, projecting that the OSW farm would be accessible (while in reality, it is not; see inset in Figure \ref{fig:intro_uncertainty}-A). Thus, the deterministic approach (mistakenly) perceives it as an optimal time to group three maintenance tasks for WT3, WT4, and WT5 to leverage this access period and concurrent low power production. With access inaccurately estimated, all three maintenance tasks end up being interrupted (See Figure \ref{fig:intro_uncertainty}-C), leaving the WTs in a non-operational status until the following day, where maintenance is ultimately resumed once access is restored.

The stochastic approach, on the other end, recognizes the forecast uncertainty (by means of the probabilistic scenarios), and hence, the risk of maintenance interruption is foreseen. Here, the model delays the maintenance until day $7$, where both wind and wave conditions are expected to be tolerable, resulting in uninterrupted maintenance for all three WTs, and a total cost reduction of $31$\% relative to the deterministic approach's schedule.

It is natural to expect that in dynamic operational environments, such as that of OSW farms, similar cases caused by the synergistic effects of parameter uncertainty occur quite often. It is therefore necessary for a maintenance scheduling approach to be able to account, not only for maintenance opportunities, but also for the uncertainties in the environmental and operational parameters that define what constitutes a maintenance opportunity in the first place. Looking at Table \ref{tab:references}, one can notice a lack of efforts that attempt to holistically consider maintenance opportunities \textit{and} the full spectrum of parameter uncertainty in a unified framework. This is the research gap addressed in this work. 

\subsection{Contributions} 
To fill this research gap, we propose the \underline{STOC}hastic \underline{H}olistic \underline{O}pportunistic \underline{S}cheduler (STOCHOS)\textemdash In Greek, STOCHOS (written as Στόχος) translates to ``target,'' or ``objective.'' The contributions of this work are listed as follows: \vspace{-0.3cm} 
\begin{enumerate}
    \item STOCHOS is an offshore-tailored maintenance optimization framework which uniquely (\textit{i}) leverages the combined maintenance opportunities arising due to accessibility, revenue, and transportation costs, and (\textit{ii}) accounts for the multi-source uncertainties of metocean conditions, asset degradation, and electricity prices. Compared to the related literature, STOCHOS presents a novel modeling framework for opportunistic maintenance optimization under uncertainty, while its rolling horizon solution algorithm is in line with the industrial practice of OSW maintenance planning. 
    \item Instead of using temporally invariant distributional assumptions about input parameters, as prevalent in the stochastic optimization literature, we generate scenarios (or trajectories) from a probabilistic forecasting framework. The trajectories 
    recognize the temporal nature of OSW data, 
    and constitute optimal inputs to the stochastic optimization model. We show the economic value of this 
    approach relative to using marginal predictive distributions or point predictions. 
    \item In terms of practical relevance, we extensively evaluate STOCHOS using real-world data from a unique geographical region in the U.S. North Atlantic where several large-scale projects are currently in-development \citep{lease2021}. We compare STOCHOS against a set of prevalent maintenance benchmarks, and show its superior performance in terms of several O\&M metrics. Our data, models, and findings can therefore provide timely and key insights to the operators of those soon-to-be-operational OSW farms.
\end{enumerate}

The remainder of the paper is organized as follows. In Section \ref{s:model}, we present the formulation of STOCHOS. Section \ref{s:data} presents the probabilistic forecasting and scenario generation approaches, followed by the solution procedure in Section \ref{s:sol}. Section \ref{s:results} presents our results and findings. We summarize our conclusions and future research directions in Section \ref{s:conclusion}. 

\section{Model Formulation} \label{s:model}
This section starts with an introduction of notation and modeling assumptions, followed by a detailed description of the mathematical model of STOCHOS. 
\subsection{Problem description and assumptions} 
\begin{enumerate}
    \item[($\mathcal{A}1$)] We define $\mathcal{I} := \{i \hspace{.1cm} | \hspace{.1cm} i \in \mathbb{Z}^+, 1 \leq i \leq N_{\mathcal{I}}\}$ as the set of offshore WTs that must undergo a maintenance task of $\tau_i$ hours (called the repair time). We focus on minor to medium tasks which typically require less than a day to complete (e.g. electrical components, grease/oil cooling liquids, sensors and controls), and comprise $\sim$\hspace{-0.05mm}$75$\% of all maintenance tasks in OSW farms \citep{carroll2016failure, dinwoodie2015reference}. 
    
    \item[($\mathcal{A}2$)] The planning horizon is divided into two sub-horizons \citep{koltsidopoulos2020data}: a day-ahead short-term horizon (STH) and a long-term horizon (LTH). The STH, $\mathcal{T} := \{t \hspace{.1cm} | \hspace{.1cm} t \in \mathbb{Z}^+, 1 \leq t \leq 24 \text{  hours}\}$, is of hourly resolution. The LTH, $\mathcal{D} := \{d \hspace{.1cm} | \hspace{.1cm} d \in \mathbb{Z}^+, 1 \leq d \leq N_{\mathcal{D}} \text{  days}\}$, has a daily resolution, starting the day after the STH up to 
    $N_{\mathcal{D}}$ days ahead. We denote LTH variables and parameters with a superscript $L$ to distinguish them from their STH counterparts. 
    
    \item[($\mathcal{A}3$)] STOCHOS is solved using sample average approximation by optimizing a stochastic MILP (to be presented next) over a random scenario subset, $\mathcal{S} := \{s \hspace{.1cm} | \hspace{.1cm} s \in \mathbb{Z}^+, 1 \leq s \leq N_{\mathcal{S}} \text{  scenarios}\}$. Hereinafter, notations for random variables, real data, point forecasts, and scenarios are denoted as follows. If $z(t)$ denotes an arbitrary random variable in the STH, then $z_t$ denotes its true (actual) value at the $t$-th hour, $\hat{z}_t$ denotes the corresponding point forecast, while ${z}_{t,s}$ is the realization of a random scenario $s$ at time $t$. Replacing $t$ by $d$ extends this notation to an LTH variable $z^L(d)$. 
    
    \item[($\mathcal{A}4$)] We assume that the wind speed, wave height, and electricity prices are uncertain during both the STH and LTH, which are denoted by $\nu(t), \eta(t), \kappa(t)$ for STH, and $\nu^L(d), \eta^L(d), \kappa^L(d)$ for LTH, respectively. For asset degradation, we assume that the OSW turbine is an integrated system with a system-level residual life (RL), denoted by 
    $\lambda^L(i) ~\forall~ i \in \mathcal{I}$, which is a turbine-specific random variable defined as the time (in days) at which the $i$th WT fails. In practice, predictions about $\lambda^L(i)$ can be obtained using a condition monitoring system that considers the impact of dynamic environmental and operational degradation parameters  \citep{bian2015degradation}. We assume that the RL is only uncertain in the LTH, since the STH, which is the day-ahead horizon, is a small interval of time relative to the typical degradation process of WTs. 
    A WT for which the realized RL is smaller than a day is bound to fail by the end of the STH (unless it is maintained). Maintenance tasks performed before the realized (true) RL are preventive (PM), otherwise they are corrective (CM). 
    
    \item[($\mathcal{A}5$)] The power produced from a WT is primarily a function of the hub-height wind speed, and is estimated using historical data \citep{ezzat2018spatio, ding2019data}. Details of the wind-to-power conversion are discussed in the supplementary materials, SM-1.
    
    \item[($\mathcal{A}6$)] WTs are accessed by crew transport vessels (CTVs) which are subject to accessibility constraints defined by wind speed and wave height safety thresholds, denoted as $\nu_{max}$ and $\eta_{max}$, respectively. Maintenance operations are subject to interruptions due to access constraints, and can be resumed once access is restored. To formally account for maintenance interruptions, we associate each maintenance task with a ``mission time,'' $A(t,i)$ which is a random variable denoting the time needed to complete a task of $\tau_i$ hours starting at time $t$. $A(t,i)$ is a function of $\tau_i$ and the
    access state denoted by $X(t) \in \{0,1\}$, which, in turn, is a function of the stochastic metocean conditions $\nu(t)$ and $\eta(t)$, and the safety thresholds $\nu_{max}$ and $\eta_{max}$. Details about the estimation of $X(t)$ and $A(t,i)$ given metocean conditions and maintenance times are provided in the supplementary materials SM-2. Note that while maintenance interruptions are allowed, longer mission times directly translate to prolonged work hours (i.e., higher crew and vessel contracting costs), and longer downtimes (i.e., higher revenue losses). 
\end{enumerate}

\subsection{The optimization model}
There exist different approaches to uncertainty modeling in the maintenance optimization literature. One common approach is to directly encode a set of assumed probability distributions about the uncertain parameters into the optimization model, which often results in nonlinear models that are typically solved using heuristic approaches \citep{shafiee2015opportunistic, wang2019optimizing, ding2012opportunistic, lu2017opportunistic, sarker2016minimizing}. Alternatively, stochastic programming uses sampling methods to integrate realizations of those probability distributions to minimize maintenance costs (or similar objectives) \citep{besnard2011stochastic, yildirim2017integrated,fallahi2022chance}. STOCHOS belongs to the second cluster, and is formulated as a two-stage stochastic mixed linear integer program (MILP). The following binary variables constitute the decision variables in STOCHOS: 

\hspace{-6mm} $m_{t,i}$ \hspace{1.3mm} : A maintenance task for WT $i$ is scheduled to start at hour $t$ in the STH.

\hspace{-6mm} $r$\hspace{8.6mm}: A vessel is rented for the STH.  

\hspace{-6mm} $m^L_{d,i,s}$ : A maintenance task is scheduled for WT $i$ at day $d$ in the LTH, in scenario $s$.

\hspace{-6mm} $r^L_{d,s}$\hspace{5mm}: A vessel is rented at day $d$ in the LTH, in scenario $s$. 

The objective, shown in (\ref{eq1s}), is to maximize the profit in the planning horizon, comprising the day-ahead profit ${l^{STH}}$, the long-term profit ${\{l^{LTH}_d\}_{d\in\mathcal{D}}}$, and a stochastic penalty term. \begin{multline}\label{eq1s}
    \max_{m_{t,i}, m^L_{d,i,s}, r, r^L_{d,s}}
    \bigg\{\overbrace{l^{STH}}^{\text{short-term profit}} + \overbrace{\sum_{d\in \mathcal{D}} l^{LTH}_d}^{\text{long-term profit}} \\ 
    - \underbrace{\frac{1}{N_S}\sum_{s\in\mathcal{S}} \bigg[ \overbrace{\sum_{i\in \mathcal{I}}(\mbox{U}_s\cdot w_{i,s} + \mbox{Y}_s\cdot b_{i,s})}^{\text{prolonged interruptions}} +\overbrace{\mbox{C}_1\cdot a_s^x + \mbox{C}_2\cdot a_s^q}^{\text{spot resource contracting}} \bigg] }_{\text{stochastic penalty term}} \bigg\}.
\end{multline}

The short-term profit, $l^{STH}$, defined as in (\ref{eq:eq2s}), is the difference between the day-ahead operating revenues and the maintenance costs. The revenue is generated from selling the power output of the $i$th WT at time $t$ and scenario $s$, denoted by $p_{t,i,s}$, at an hourly market price $\kappa_{t,s}$. The maintenance costs comprise four components: (1) the repair costs which are dictated by the PM and CM cost coefficients, K $(\$/\mbox{task})$ and $\Phi~(\$/\mbox{task})$, respectively, (2) the daily vessel rental cost with a daily rate of $\Omega~(\$/\mbox{day})$, (3) crew-related costs charged at the crew hourly cost rate $\Psi~(\$/\mbox{h})$ where $x_{t,i,s} \in \{0,1\}$ denotes whether a crew is assigned to the $i$th WT at time $t$ and scenario $s$, and (4) overtime costs charged at the crew overtime cost rate $\mbox{Q}~(\$/\mbox{h})$ where $q_s \in \mathbb{Z}^+$ denotes the overtime hours worked in the STH. Binary parameter $\zeta_i$ indicates the day-ahead operational status of the $i$-th WT, wherein $\zeta_i = 0$ indicates a failed WT needing a CM. The parameter $\rho_i \in \{0,1\}$ denotes whether a maintenance task has been initiated in a previous day but is yet to be completed (when $\rho_i = 1$, repair cost goes to zero as it has already been accounted for in a previous day). 
\begin{dmath}\label{eq:eq2s}
    l^{STH}=-\sum_{i \in \mathcal{I}} \sum_{t \in \mathcal{T}} \overbrace{(1-\rho_i)\cdot[\zeta_i\cdot\mbox{K}+(1-\zeta_i)\cdot\Phi) \cdot m_{t,i}}^{\text{repair cost}} - \overbrace{\Omega \cdot r}^{\text{vessel cost}}+  \frac{1}{N_S}\sum_{s \in \mathcal{S}}\left[\sum_{i \in \mathcal{I}} \sum_{t \in \mathcal{T}}(\underbrace{\kappa_{t,s} \cdot  p_{t,i,s}}_{\text{operating revenue}}  - \underbrace{\Psi \cdot x_{t,i,s}}_{\text{crew cost}}) - \underbrace{\mbox{Q} \cdot q_s}_{\text{overtime cost}} \right].
\end{dmath}

The long-term profit, in (\ref{eq:eq3s}), is similarly calculated for daily intervals. The crew work hours are calculated as the the mission time $A_{d,i,s}^L$ of the tasks scheduled at that day, $m_{d,i,s}^L$. 

\begin{dmath} \label{eq:eq3s}
 l^{LTH}_d = \frac{1}{N_S}\sum_{s\in\mathcal{S}}\left\{\sum_{i \in \mathcal{I}} \bigg[ \overbrace{\kappa_{d,s}^L\cdot p_{d,i,s}^L}^{\text{operating revenue}} - \overbrace{(1-\rho_i)\cdot[\zeta_{d,i,s}^L\cdot\mbox{K}+(1-\zeta_{d,i,s}^L)\cdot\Phi] \cdot m_{d,i,s}^L}^{\text{repair cost}} - \underbrace{\Psi\cdot {A_{d,i,s}^L} \cdot m_{d,i,s}^L}_{\text{crew cost}} \bigg] - \underbrace{\Omega \cdot r_{d,s}^L}_{\text{vessel cost}} - \underbrace{\mbox{Q} \cdot q_{d,s}^L}_{\text{overtime cost}}\right\} \quad {\forall ~ d \in \mathcal{D}}.
\end{dmath}

The last term in (\ref{eq1s}) is a stochastic penalty which entails two sub-terms. The first term, which we call ``prolonged interruptions,'' represents the cost of maintenance actions that started in the STH, but were not completed, and hence, must be resumed in the LTH (The WT remains non-operational when such event takes place). In that term, $\mbox{U}_s \cdot w_{i,s}$ represent the cost incurred \textit{until} the maintenance is restarted (pre-maintenance), while $\mbox{Y}_s \cdot b_{i,s}$ is the cost incurred \textit{while} the maintenance is ongoing (during maintenance). In specific, $w_{i,s} \in \{0,1\}$ denotes the occurrence of the interruption event, $\mbox{U}_s$ is the associated pre-maintenance cost, $b_{i,s} \in \mathbb{Z}^+$ denotes the remaining maintenance time that has to be completed in the LTH, and $\mbox{Y}_s$ is the associated cost incurred during the maintenance. The second sub-term penalizes the (unlikely) case when the maximum budget for crew (\ref{eq:workhourss}) or overtime resources (\ref{eq:crew2s}) are exceeded. We assign fairly large cost coefficients, $\mbox{C}_1$ and $\mbox{C}_2$, for such exceedances to reflect the unavoidable practice of on-the-spot contracting of additional crew, equipment, or other immediate maintenance resources, when necessary. In our experiments, those exceedances occur less than $0.09\%$ of the time.

\hspace{-7mm} \underline{Maintenance Constraints:} The equality in (\ref{eq:maints}) forces a maintenance task to be scheduled either in the STH or in the LTH.
\begin{equation} \label{eq:maints}
\sum_{t \in \mathcal{T}}m_{t,i}+\sum_{d \in \mathcal{D}}m_{d,i,s}^L = 1 \quad \forall  ~ i \in \mathcal{I}, s \in \mathcal{S}.
\end{equation}

A maintenance task in the STH can only be initiated after the time of first light, $t_R$, and before the time of last sunlight, $t_D$, as expressed in (\ref{eq:sunrises}) and (\ref{eq:sundowns}), respectively.
\begin{equation} \label{eq:sunrises}
{m_{t,i} \leq \frac{t}{t_R} \quad \forall ~ t \in \mathcal{T}, i \in \mathcal{I}}.
\end{equation}
\begin{equation} \label{eq:sundowns}
{m_{t,i} \leq \frac{t_D}{t} \quad \forall ~ t \in \mathcal{T}, i \in \mathcal{I}}.
\end{equation}

Once a maintenance task is initiated at time $t$ for the $i$th WT, then it would be under maintenance for a period of time computed as the minimum between the remaining time in the STH, which is $24-t$, and the mission time $A_{t,i,s}$. This is expressed in (\ref{eq:downs}), where the binary variable $u_{\tilde{t},i,s}$ denotes a turbine under maintenance at time $\tilde{t}$.  
\begin{equation} \label{eq:downs}
{\sum_{\tilde{t}=t}^{\min\{24, t+A_{t,i,s}\}} u_{\tilde{t},i,s} \geq \min\{24-t, A_{t,i,s}\} \cdot m_{t,i} \quad \forall t \in \mathcal{T}, i \in \mathcal{I}, s\in \mathcal{S}}.
\end{equation}

If that maintenance task is not completed within the STH, then $b_{i,s} \in \mathbb{Z}^+$, as shown in (\ref{eq:incomplete2s}), denotes the remaining time required to complete it in the LTH, while $w_{i,s} \in \{0,1\}$, as shown in (\ref{eq:incompletes}), denotes the occurrence of such event and is only set to $1$ once $b_{i,s} > 0$, as it is multiplied by M, an arbitrary large number. This is the case where the ``prolonged interruption'' penalty term in (\ref{eq1s}) is activated.
\begin{equation} \label{eq:incomplete2s}
{b_{i,s} \geq \sum_{t\in \mathcal{T}}m_{t,i}\cdot [A_{t,i,s}- 24 +t ]^+
 \quad \forall i \in \mathcal{I}, s\in \mathcal{S}}.
\end{equation}

\begin{equation} \label{eq:incompletes}
{b_{i,s} \leq \mbox{M} \cdot w_{i,s}
 \quad \forall i \in \mathcal{I}, s\in \mathcal{S}}.
\end{equation}

The maintenance crew, once dispatched, is occupied until the maintenance is completed or their shift ends by the time of last sunlight, $t_D$, as shown in (\ref{eq:crews}). An upper bound on the number of available maintenance crews is set to $\mbox{B}$ (crews), as shown in (\ref{eq:crew2s}).
The auxiliary variable $a_s^x \in \mathbb{Z}^+$ is added to the right-hand-side (RHS) of (\ref{eq:crew2s}) to 
account for the (unlikely) case in which the crew budget limit is exceeded, but almost always remains zero since it is heavily penalized by $\mbox{C}_1$ in (\ref{eq1s}). In practice, this translates to situations where additional crew or equipment is contracted on-the-spot. 
\begin{equation} \label{eq:crews}
{x_{t,i,s} \geq u_{t,i,s} - \frac{t}{t_D}
 \quad \forall t \in \mathcal{T}, i \in \mathcal{I}, s\in \mathcal{S}}.
\end{equation}
\begin{equation} \label{eq:crew2s}
\sum_{i \in \mathcal{I}} x_{t,i,s} \leq \mbox{B} + a_s^x \quad \forall t \in \mathcal{T}, s\in \mathcal{S}.
\end{equation}

\hspace{-6mm}\underline{Turbine Availability Constraints:} A failed WT (i.e., one which has not been maintained at or before its RL) remains unavailable until a maintenance action is completed. For the STH case, availability can be expressed by the binary variable $y_{t,i,s}$, as in (\ref{eq:availabilitys}).
\begin{dmath} \label{eq:availabilitys}
y_{t,i,s} \leq \overbrace{\zeta_i\cdot(1-\rho_i)}^{\text{WT operational status}} + \overbrace{\frac{24 \cdot \sum_{\tilde{t} \in \mathcal{T}}m_{\tilde{t},i}- \sum_{\tilde{t} \in \mathcal{T}}(\tilde{t} \cdot m_{\tilde{t},i})}{24-t+g}}^{\text{availability restored after maintenance}}
{\quad \forall t \in \mathcal{T}, i \in \mathcal{I}, s\in \mathcal{S} }.
\end{dmath}

The first term of the RHS in (\ref{eq:availabilitys}) denotes whether the turbine is in a failed state at the beginning of the STH, or if a maintenance task has been initiated in a previous day. In case the RL is reached, then the turbine fails ($\zeta_i=0$), and can only return to its former operational status once a CM action is performed. 
This is enforced by the second term in the RHS; if no CM is scheduled, then $m_{\tilde{t},i}=0$ and the term drops to 0. In contrast, if $m_{\tilde{t},i}=1$, the term is greater than $1$ after the time of maintenance, $\tilde{t}$. An arbitrary small number $g$ avoids division by zero. A similar constraint for the LTH is shown in~(\ref{eq:availability2s}).
\begin{dmath} \label{eq:availability2s}
y_{d,i,s}^L \leq {\zeta_{d,i,s}^L\cdot (1-\rho_i)}+\frac{N_{\mathcal{D}}- \sum_{d \in \mathcal{D}}(d \cdot m_{d,i,s}^L)}{N_{\mathcal{D}}-d+g} \hspace{0.3cm} {\forall d \in \mathcal{D}, i \in \mathcal{I}, s\in \mathcal{S} }.
\end{dmath}

A WT under maintenance remains unavailable until the task is completed, as in (\ref{eq:maintts}).
\begin{equation} \label{eq:maintts}
   y_{t,i,s} \leq 1 - {u_{t,i,s}} \quad \forall t \in \mathcal{T}, i \in \mathcal{I}, s\in \mathcal{S}.
\end{equation}

\hspace{-6mm}\underline{Vessel Rental Constraints:} Vessels are rented daily only if a maintenance task has been scheduled at that day. This is enforced via (\ref{eq:vessel1s})-(\ref{eq:vessel2s}), for the STH and LTH, respectively.
\begin{equation} \label{eq:vessel1s}
    \mbox{M}\cdot r \geq \sum_{t \in \mathcal{T}} \sum_{i \in \mathcal{I}} m_{t,i}. 
\end{equation}
\begin{equation}\label{eq:vessel2s}
  \mbox{M} \cdot r_{d,s}^L \geq \sum_{i \in \mathcal{I}}m_{d,i,s}^L \quad \forall d \in \mathcal{D}, s\in \mathcal{S}. 
\end{equation}

\hspace{-6mm}\underline{Overtime Constraints:} The total work hours, as shown in (\ref{eq:workhourss}), cannot exceed $\mbox{W}$ (hour/crew). Otherwise, overtime, tracked by integer variable $q_s$, are incurred and compensated at a higher rate determined by $\mbox{Q}$ (\$/hour). An upper bound of $\mbox{H}$ (in hours) limits the number of overtime hours in both the STH and LTH, as in (\ref{eq:overtime1s}) and (\ref{eq:overtime2s}). Similar to $a^x_s$ in (\ref{eq:crew2s}), $a_s^q$ in (\ref{eq:workhourss}) accounts for  the on-spot contracting of additional crew, at a large cost of $\mbox{C}_2$.

A task initiated but not finished in the STH, is prioritized in the first day of the LTH ($d = 1$). This is expressed in (\ref{eq:workhours2s}) where $A_{d=1,i,s}^L \cdot m_{d=1,i,s}^L + b_{i,s}$ denotes the total work hours in the first day of the LTH (the sum of the mission times and the remaining maintenance time of unfinished tasks). A similar constraint in (\ref{eq:workhours3s}) is imposed for the remaining days of the LTH, i.e. for $d \in \{2, ..., N_D\}$.
\begin{equation} \label{eq:workhourss}
\sum_{t \in \mathcal{T}, i \in \mathcal{I}} x_{t,i,s} \leq \mbox{B} \cdot \mbox{W}+q_s+a_s^q \quad \forall s\in \mathcal{S}.
\end{equation}
\begin{equation} \label{eq:workhours2s}
\sum_{i \in \mathcal{I}} \left[A_{1,i,s}^L \cdot m_{1,i,s}^L + b_{i,s}\right] \leq \mbox{B} \cdot \mbox{W} + q_{1,s}^L \quad \forall 
s\in \mathcal{S}.
\end{equation}
\begin{equation} \label{eq:workhours3s}
\sum_{i \in \mathcal{I}} \left[A_{d,i,s}^L \cdot m_{d,i,s}^L\right] \leq \mbox{B} \cdot \mbox{W} + q_{d,s}^L \quad \forall 
d \in \{2, ..., N_{\mathcal{D}}\}, s\in \mathcal{S}.
\end{equation}
\begin{equation} \label{eq:overtime1s}
q_s \leq \mbox{H} \quad \forall s\in \mathcal{S}.
\end{equation}
\begin{equation} \label{eq:overtime2s}
q_{d,s}^L \leq \mbox{H} \quad \forall d \in \mathcal{D}, s\in \mathcal{S}.
\end{equation}

\hspace{-6mm}\underline{Electricity Generation Constraints:} The hourly power output, $p_{t,i,s}$, is computed as a fraction $f_{t,i,s} \in [0,1]$ of the turbine's rated capacity $\mbox{R}$ (MW), multiplied by the turbine's availability $y_{t,i,s}$, as shown in (\ref{eq:eqnoms}). 
When the WT is in a failed state, or under maintenance, $p_{t,i,s} = y_{t,i,s} = 0$, and the operator forfeits the associated revenue. The fraction $f_{t,i,s}$ is called \textit{the normalized power level}. Full details of estimating $f_{t,i,s}$ and $f_{d,i,s}^L$ given wind speed conditions are shown in the supplementary materials, SM-1. \begin{equation}\label{eq:eqnoms}
    p_{t,i,s}\leq\mbox{R} \cdot f_{t,i,s} \cdot y_{t,i,s} \quad \forall t \in \mathcal{T}, i \in \mathcal{I}, s\in \mathcal{S}.
\end{equation}

Likewise, the daily power output, $p_{d,i,s}^L$, is defined in (\ref{eq:nom2s}), wherein $f_{d,i,s}^L \in [0, 1]$ is a function of the daily average wind speed.  
The term $m_{d,i}^L \cdot\zeta_{d,i,s}^L\cdot A_{d,i,s}^L/24$, accounts for the production losses if a preventive maintenance has been scheduled on day $d$.
\begin{equation} \label{eq:nom2s}
p_{d,i,s}^L \leq 24 \cdot \mbox{R} \cdot f_{d,i,s}^L \cdot \left(y_{d,i,s}^L - m_{d,i,s}^L \cdot \frac{A_{d,i,s}^L\cdot\zeta_{d,i,s}^L}{24}  \right) \quad \forall d \in \mathcal{D}, i \in \mathcal{I}, s\in \mathcal{S}.
\end{equation}

Curtailing wind power production is common in grid-connected wind farms. 
This is accounted for by introducing the parameter $\mbox{C}_{t,s}\in [0,1]$ in (\ref{eq:curtailment}) which defines the fraction of the farm-level power output that can be injected to the grid at each hourly interval $t$. $C_{t,s} = 1$ denotes no curtailment\textemdash i.e., all power produced is eventually sold.  
\begin{equation} \label{eq:curtailment}
\sum_{i \in \mathcal{I}} p_{t,i,s} \leq \sum_{i \in \mathcal{I}}f_{t,i,s} \cdot \mbox{R} \cdot \mbox{C}_{t,s} \quad \forall t \in \mathcal{T}, s\in \mathcal{S}.
\end{equation}

\section{Probabilistic Forecasting and Scenario Generation} \label{s:data}
We first introduce the data used in this work and its relevance to the U.S. OSW industry, then present our probabilistic forecasting and scenario generation framework.
\subsection{Data description}
This work is motivated by the ongoing large-scale OSW developments in the US North Atlantic, and in particular, the NY/NJ Bight, 
where several Gigawatt-scale OSW projects are in-development \citep{lease2021}. We use real-world OSW data and meteorological forecasts from this region, in 
proximity to those soon-to-be-operational OSW projects. 

\begin{figure}[h]
    \centering
    \includegraphics[trim = {0cm, .25cm, 0, 0cm}, width=.65\linewidth]{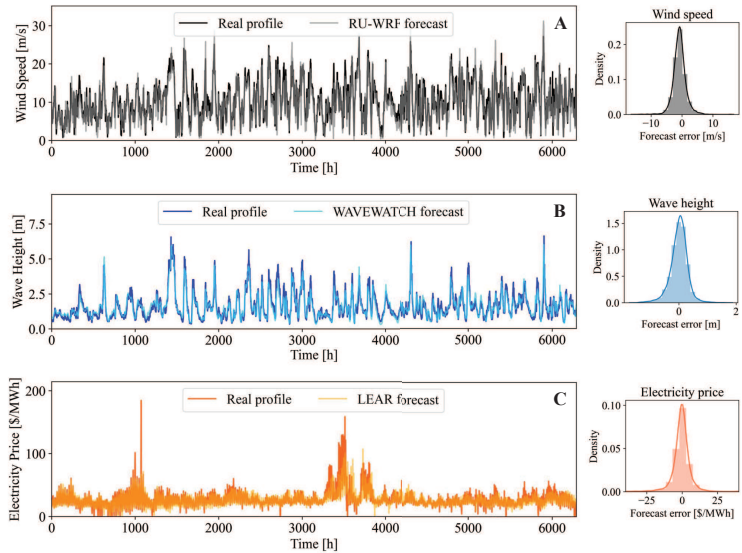}
    \caption{Time series of the data and forecasts of wind speed (A), wave height (B) and electricity price (C). Histograms of the forecast residuals are shown to the right.} 
    \label{fig:ts_data}
\end{figure}

\hspace{-6.3mm}\underline{Metocean Data and Forecasts:} We use wind speed and wave height measurements collected by the E05 Hudson North buoy between August 2019 and May 2020. 
This is one of two floating LiDAR buoys recently deployed by the NY State Energy Research \& Development Authority (NYSERDA) \citep{nyserda2019} in the NY/NJ Bight. Co-located with the measurements are numerical weather predictions (NWPs) of wind speed and wave height. The wind speed NWPs come from RU-WRF, a meso-scale weather model developed by the Rutgers University Center For Ocean Observing Leadership (RUCOOL) to capture the unique physical phenomena in this geographical region 
\citep{optis2020validation}. 
Wave height NWPs come from WAVEWATCH III, a numerical wave model maintained by the National Oceanic and Atmospheric Administration (NOAA). 
The metocean data and their NWPs are shown in Figures \ref{fig:ts_data}-A and \ref{fig:ts_data}-B, respectively. 

\hspace{-6.5mm}\underline{Electricity Price Data and Forecasts:} Real day-ahead electricity prices are obtained from PJM's Data Miner III \citep{PJM}, at node COMED (node id: $33092371$). Point forecasts for the same node are obtained by a Lasso Estimated Auto-Regressive (LEAR) model proposed in \cite{lago2021forecasting}. The time series of the real electricity prices, the LEAR predictions, and the associated error histograms are shown in Figure \ref{fig:ts_data}-C.

\hspace{-6.5mm}\underline{Asset Degradation Data and Forecasts:} Unlike the above-listed parameters (metocean conditions and electricity prices), actual turbine degradation data requires access to O\&M records. Currently, there are no operational OSW farms in this region. So, we assume a set of observed RLs, $\{\lambda_{i}\}_{i \in \mathcal{I}}$, which are unknown to the planner, who only has access to a set of (imperfect) RL predictions $\{\hat{\lambda}_i\}_{i \in \mathcal{I}}$ provided by a condition monitoring system. In this study, true and predicted RLs are carefully selected in light of the summary statistics derived in previous studies on offshore WT reliability analyses \citep{carroll2016failure}, thus ensuring a faithful replication of real-world WT failures (More details in Section 5.1).

\subsection{Uncertainty modeling and scenario generation}
\begin{figure}[h]
    \centering
    \includegraphics[width=0.6\linewidth]{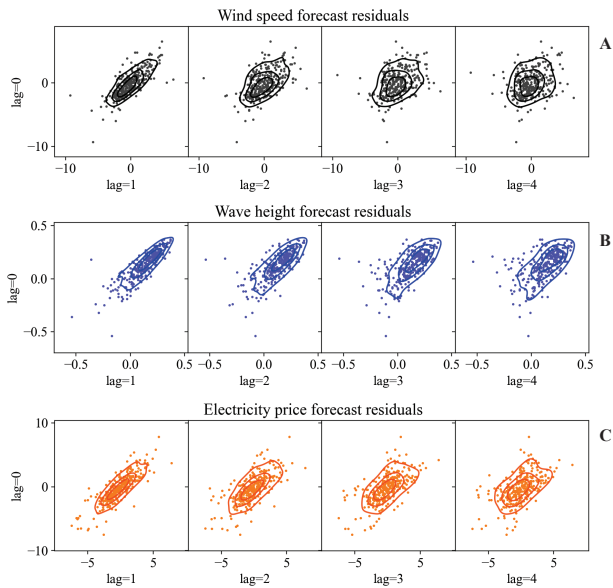}
    \caption{Density plots of the point forecast residuals versus their lagged versions for wind speed (A), wave height (B) and electricity price (C) time series forecasts.} 
    \label{fig:ts_residuals}
\end{figure}

Modeling uncertainty in scenario-based stochastic optimization requires probabilistic characterizations of the input parameters. A prevalent approach in the literature is to impose a distributional assumption on the marginal distribution of the forecast residuals at each lead time. Albeit convenient, 
this approach is sub-optimal when temporal dependencies exist, as in the case of OSW data. 
Figure \ref{fig:ts_residuals} depicts the density of the forecast residuals at four time lags for wind speed (top), wave height (middle), and electricity price (bottom).       

Instead, a more powerful approach is to fully describe the density of the temporal process governing each input parameter. Scenarios generated from such approach would preserve the temporal dependence and constitute the optimal input to a stochastic program \citep{pinson2013wind}. Specifically, we define $z(t)$ as the random variable denoting either the wind speed $\nu(t)$, wave height $\eta(t)$, or electricity price $\kappa(t)$ (we treat the RL predictions differently as discussed at the end of this section). We also denote by $\mathbf{z} = [z_1, ..., z_{tc}]^T$ the set of historical measurements of $z(t)$ up to the current time $t_c$. Similarly, $\hat{\mathbf{z}} = [\hat{z}_1, ..., \hat{z}_{tc}]^T$ are the raw point predictions available to the planner (RU-WRF for $\nu(t)$, WAVEWATCH III for $\eta(t)$, and LEAR for $\kappa(t)$). Then, we propose the following probabilistic forecasting framework, comprising three independent terms: \begin{equation}\label{eq:fm}
    z(t) = \mu(t) + \omega(t) + \epsilon(t),
\end{equation}
where $\mu(t)$ is a deterministic mean function defined as the raw point predictions available to the maintenance planner (that is, $\mu(t) = \hat{z}_t ~\forall~ t$). The term $\omega(t)$ is a zero-mean temporal Gaussian Process (GP) with its pairwise covariance denoted by $\sigma_{t,t'} = Cov\{\omega(t), \omega(t')\}, \forall t, t'$, while $\epsilon(t)$ is the zero-mean Gaussian noise, such that $\pmb{\epsilon} = [\epsilon_1, ..., \epsilon_{t_c}]^T \sim \mathcal{N}(0,\delta \mathbf{I})$, where $\mathbf{I}$ is the $t_c \times 1$ identify matrix, and $\delta$ is the noise variance. 

Let us use $\mathbf{C}$ to denote the $t_c$ × $t_c$ covariance matrix whose $(t,t')$-th entry is defined as $\sigma_{t,t'} + \delta \mathbb{I}({t=t'})$, where $\mathbb{I}(\cdot)$ is the indicator function. Those entries are estimated using a stationary kernel $C(\cdot,\cdot)$ which encodes the temporal dependence. Several mathematically permissible choices for $C(\cdot, \cdot)$ are possible and we use the squared exponential kernel, 
for which the hyperparameters are estimated by maximizing the log-likelihood of the GP. 
Using the estimated parameters, 
we can obtain the estimated covariance matrix, $\hat{\mathbf{C}}$. 

Putting together all the pieces, one can fully characterize the predictive distribution of the random variable $z^*(t_c+h)$ which denotes the forecast of $z(t_c + h)$. By virtue of the GP, this distribution is Gaussian with its mean and variance as in (\ref{eq:gpmean}) and (\ref{eq:gpvar}), respectively. 
\begin{equation}\label{eq:gpmean}
    {\bar{z}}_{t_c+h} = \mathbb{E}[z^*(t_c+h)|\mathbf{z},\hat{\mathbf{z}}] = 
    \hat{z}_{t_c+h} + 
    \hat{\mathbf{k}}^T \hat{\mathbf{C}}^{-1} (\mathbf{z} - \hat{\mathbf{z}}),
\end{equation}
\begin{equation}\label{eq:gpvar}
\bar{\sigma}^2_{t_c+h} = \mathbb{V}[z^*(t_c+h)|\mathbf{z},\hat{\mathbf{z}}] = \hat{\alpha} - \hat{\mathbf{k}}^T \hat{\mathbf{C}}^{-1} \hat{\mathbf{k}},
\end{equation}
where $h \in \{1, ..., H\}$ is the forecast lead time, $\hat{z}_{t_c +h}$ is the raw forecast (e.g. from RU-WRF for wind speed) at $t_c + h$.
The $t_c \times 1$ vector $\hat{\mathbf{k}}$ contains the pairwise covariances between $[1, ..., t_c]^T$ and $t_c + h$, computed using $C(\cdot,\cdot)$. A similar model to (\ref{eq:fm}) is developed for the LTH (in daily resolution), and similar expressions to those in (\ref{eq:gpmean}) and (\ref{eq:gpvar}) are derived. 

The advantages of this probabilistic framework are two-fold. First is its ability to effectively account for the temporal dependencies in OSW data. This is obvious in how the predictive mean in (\ref{eq:gpmean}) depends on the correlation between the target forecast and the historical observations. This is also apparent in how the predictive variance in (\ref{eq:gpvar}) is reduced from the marginal variance when information about strongly correlated values in the process history is available. This is in contrast to distributional assumptions which treat each lead time individually, thus overlooking the value added by the temporal dependence. 

The second advantage is our ability to draw random temporal trajectories (or scenarios) by sampling from the joint multivariate Gaussian distribution of the forecasts, as shown in (\ref{eq:mvn}). The resulting trajectories, which naturally encode the temporal dependence in them, are optimal inputs to the stochastic program, as opposed to marginal predictive densities. 
\begin{equation}\label{eq:mvn}
    {\mathbf{z}^*} = [{z}^*(t_c+1), ..., {z}^*(t_c+H)]^T \sim \mathcal{N}(\bar{\mathbf{z}}, \bar{\mathbf{C}}),
\end{equation}
where $\bar{\mathbf{z}} = [\bar{z}_{t_c + 1}, ..., \bar{z}_{t_c + H}]^T$ is the $H \times 1$ vector of the GP predictive means, and $\bar{\mathbf{C}}$ is the correspondent $H \times H$ predictive covariance matrix. 
\begin{figure}[h]
    \centering
        \includegraphics[trim = 0 5.5cm 0 5.5cm, width = .72\linewidth]{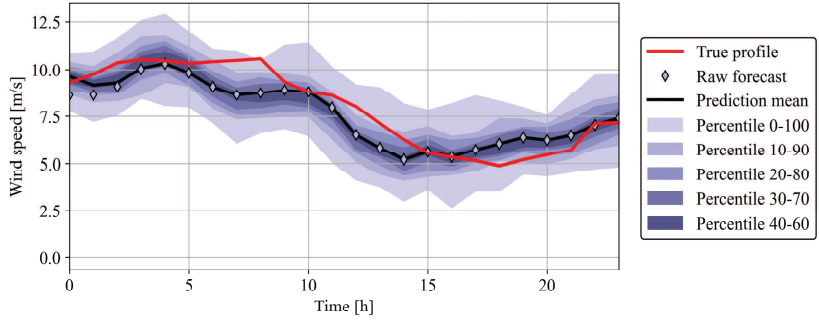}
    \caption{Fan charts of $1,000$ random scenarios (trajectories) on 09/21/2019.} 
    \label{fig:gp_regression}
\end{figure}

Figure \ref{fig:gp_regression} shows a fan chart of $1,000$ trajectories of wind speeds drawn randomly on a select day using (\ref{eq:mvn}). Similar trajectories can be drawn for the wave height and electricity prices. While the above two advantages are largely acknowledged in the wind forecasting literature \citep{pinson2013wind, ezzat2019spatio}, limited research has bridged such probabilistic forecasting models with the stochastic maintenance optimization realm. One of the key outcomes of our analyses in this paper is to demonstrate the ``economic value'' realized by an effective uncertainty modeling of temporal dependencies in input parameters. 

For turbine RL estimates, we assume a simpler probabilistic model where the prediction error from the $i$th WT is assumed to follow a Weibull distribution with $\hat{\lambda}_i$ and $\xi^\lambda_i$ as scale and shape parameters, respectively, such that the forecast variable $\lambda^*_i \sim \mbox{Weibull}(\hat{\lambda}_i,\xi^\lambda)$. The predictive mean of $\lambda^*_i$ is denoted by ${\bar{\lambda}}_i = \hat{\lambda}_i \Gamma(1+\frac{1}{\xi_i^\lambda})$.

\section{Solution Procedure and Computational Efficiency} \label{s:sol}
We solve STOCHOS using the Gurobi 9.0 solver with a Python interface in a server with two $14$-core Intel Xeon CPUs with a base frequency of $2.60$ GHz and $128$ GB of memory, for a relative optimality gap of $0.1\%$. Table \ref{tab:sol_times} shows the solution time (in seconds) as a function of the number of scenarios ($N_s = 25, 50, 100,$ and $200$) and the number of WTs ($N_{\mathcal{I}}=5, 15$ and $50$). Even with $200$ scenarios and $50$ WTs, STOCHOS is solved in under $45$ minutes. In practice, wind farm operators only need to run STOCHOS once per day, so they can use $200$ scenarios (or even more), although our experiments suggest
that the solution quality does not significantly change beyond $50$ scenarios.
\begin{table}[]
    \caption{Average solution time in seconds, for different number of scenarios ($N_{\mathcal{S}}$) and number of wind turbines ($N_{\mathcal{I}}$), using an optimality gap of $0.1$\%. }
    \vspace{+0.1cm}
    \label{tab:sol_times}
    \renewcommand{\arraystretch}{.6}
    \centering
    \begin{tabular}{| c | c | C{2.4cm} | C{2.4cm} | C{2.4cm} | C{2.4cm}|}
    \hline
     \multicolumn{2}{|c|}{
     \multirow{2}{*}{}
     } & \multicolumn{4}{c|}{ $\mathbf{N_{\mathcal{S}}}$}  \\
     \cline{3-6}
     \multicolumn{2}{|c|}{} &  $\mathbf{25}$ & $\mathbf{50}$ & $\mathbf{100}$ & $\mathbf{200}$ \\
    \hline
     
     \multirow{3}{*}{$\mathbf{N_{\mathcal{I}}}$} & $\mathbf{5}$  & $7.1$  & $14.3$  & $23.3$  & $96.9$  \\
     \cline{2-6}
     & $\mathbf{15}$ &  $24.6$  & $92.5$  & $157.2$  & $592.6$  \\
     \cline{2-6}
     & $\mathbf{50}$ &  $199.1$  & $631.1$  & $1,861.6$  & $2,531.4$  \\
    
    \hline
    \end{tabular}
    \vspace{-0.2cm}
\end{table}




In actual operations, STOCHOS would be solved at the end of each working day to inform the day-ahead maintenance plan \citep{koltsidopoulos2020data}. The rolling horizon procedure presented in Algorithm 1, explains the details of such implementation, starting from parameter and data input, to training the probabilistic forecasting models of Section 3.2, followed by the scenario generation step, in which $N_s$ trajectories for the wind speed, wave height, and electricity price are generated, together with $N_s$ turbine-specific RL prediction samples. Those scenarios are then used to determine scenario-specific power output predictions, accessibility estimations, and projected mission times. STOCHOS is then solved by minimizing the objective function in (\ref{eq1s}), returning an hourly day-ahead maintenance schedule ($m_{t,i}$ and $r$), and a daily schedule for the long-term horizon ($m_{d,i,s}^L$ and $r_{d,s}^L$). Once a solution has been obtained, the planning horizon is shifted by one day into the future. We call this as a single optimization ``roll,'' and the whole process is repeated until all maintenance tasks are completed. Finally, all STH schedules from all rolls are patched representing the executed hourly maintenance schedule for the whole planning horizon. This is the hourly schedule that is implemented by following STOCHOS' solutions, and hence, is used to assess its performance relative to other benchmarks. 
\begin{algorithm}[]
\label{alg: alg1}
\caption{A rolling-based solution procedure for STOCHOS}
\begin{algorithmic}[1]
\STATE \textit{Input} set dimensions $N_{\mathcal{I}}, N_{\mathcal{D}}, N_{\mathcal{S}}$ $\rightarrow$ \texttt{\small{\# of WTs, \# of days, and \# of scenarios}}
\vspace{-0.2cm}
\STATE \textit{Input} parameters $\mbox{K}, \Phi, \Psi, \Omega, \mbox{Q}, \mbox{R}, \mbox{B}, \mbox{W}, \tau_i, \nu_{max}, \eta_{max}$ $\rightarrow$ \texttt{\small{operational parameters}}
\vspace{-0.2cm}
\STATE \textit{Initialize}  $\theta_i=1 \quad \forall i\in\mathcal{I}$ $\rightarrow$ \texttt{\small{parameter denoting maintenance requirement}}
\vspace{-0.2cm}
\STATE \textit{Initialize}  $\rho_i=0, \zeta_i = 1 \quad \forall i\in\mathcal{I}$ $\rightarrow$ \texttt{\small{parameters denoting WT operational status}}
\vspace{-0.2cm}
\STATE \textit{Set} the roll counter $j = 0$ 
\vspace{-0.2cm}
\WHILE{$\sum_{i\in\mathcal{I}}\theta_i>0$}
\vspace{-0.2cm}
\STATE \textit{Set} $\mathcal{T}=\{24\cdot j,..., 24\cdot(j+1)\}$; $\mathcal{D}=\{j+1,..., N_{\mathcal{D}}+j\}$
\vspace{-0.2cm}
\STATE \textit{Input} observations and predictions for $\nu(t)$, $\eta(t)$, $\kappa(t)$, and $\lambda(i) ~\forall i$. \vspace{-0.2cm}
\STATE \textit{Train} the STH probabilistic models of Section 3.2 for $\nu(t)$, $\eta(t)$, and $\kappa(t)$. \vspace{-0.2cm}
\STATE \textit{Train} the LTH probabilistic models of Section 3.2 for $\nu^L(d)$, $\eta^L(d)$, $\kappa^L(d)$, $\lambda^L(i) ~\forall i$. \vspace{-0.2cm}
\STATE \textit{Sample} $N_{\mathcal{S}}$ scenarios $\{{\nu}_{t,s}\}_{t\in \mathcal{T}}^{ s\in \mathcal{S} }$, 
$\{{\eta}_{t,s}\}_{t\in \mathcal{T}}^{ s\in \mathcal{S}}$, 
and $\{{\kappa}_{t,s}\}_{t\in \mathcal{T}}^{ s\in \mathcal{S} }$ 
$\rightarrow$ \texttt{\small{STH scenarios for wind speed, wave height \& price}} \vspace{-0.2cm}
\STATE \textit{Sample} $N_{\mathcal{S}}$ scenarios $\{{\nu}_{d,s}^L\}_{d\in \mathcal{D}}^{ s\in \mathcal{S} }$, 
$\{{\eta}_{d,s}^L\}_{d\in \mathcal{D}}^{ s\in \mathcal{S} }$, 
and $\{{\kappa}_{d,s}^L\}_{d\in \mathcal{D}}^{ s\in \mathcal{S} }$ 
$\rightarrow$ \texttt{\small{LTH scenarios for wind speed, wave height \& price}} \vspace{-0.2cm}
\STATE \textit{Sample} $N_{\mathcal{S}}$ scenarios $\{{\lambda^L}_{i,s}\}_{i\in \mathcal{I}, s\in \mathcal{S}}$ 
$\rightarrow$ \texttt{\small{scenarios for turbine RL}} \vspace{-0.2cm}  
\STATE \textit{Evaluate} $\{{f}_{t,s}\}_{t \in \mathcal{T}}^{s \in \mathcal{S}}$ and $\{{f}_{d,s}^L\}_{d \in \mathcal{D}}^{s \in \mathcal{S}}$ $\rightarrow$ \texttt{\small{Speed-to-power conversion; SM-1}} \vspace{-0.2cm}
\STATE \textit{Evaluate} $\{{A}_{t,i,s}\}_{t \in \mathcal{T}}^{s \in \mathcal{S}}$ and $\{{A}_{d,i,s}^L\}_{d \in \mathcal{D}}^{s \in \mathcal{S}}$ $\rightarrow$ \texttt{\small{Computing STH and LTH mission times; SM-2}} \vspace{-0.75cm}
\STATE \textit{Solve} STOCHOS for $N_{\mathcal{S}}$ scenarios
\vspace{-0.2cm}
\STATE \textit{Return} the decisions for the $j$th roll: $\mathcal{S}^{STH}_j = \{m_{t,i}, r\}$ and $\mathcal{S}^{LTH}_j = \{m_{d,i,s}^L, r_{d,s}^L\}$. 
\vspace{-0.2cm}
\STATE \textit{Evaluate} $\mathcal{S}^{STH}_j$ under real operational parameters
\vspace{-0.2cm}
\FOR{$i\in \mathcal{I}$}
\vspace{-0.2cm}
\IF{$\sum_{t\in\mathcal{T}}m_{t,i}>0$} 
\vspace{-0.2cm}
\IF{$b_i>0$}
\vspace{-0.2cm}
\STATE \textit{Set} $\rho \leftarrow 1$; $\tau_i \leftarrow b_i$ (\texttt{\small{maintenance resumed if unfinished in the STH}})
\vspace{-0.2cm}
\ELSE
\vspace{-0.2cm}
\STATE \textit{Set} $\theta_i \leftarrow 0$ (\texttt{\small{maintenance no longer required if completed in the STH}})

\vspace{-0.2cm}
\ENDIF
\vspace{-0.2cm}
\ENDIF
\vspace{-0.2cm}
\ENDFOR
\vspace{-0.2cm}
\STATE \textit{Update} roll counter $j= j+1$ \vspace{-0.2cm}
\STATE \textit{Update} RLs \& WT status $\lambda_i=(\lambda_i-1, 0)^+$ ; $\hat{\lambda}_i=(\hat{\lambda}_i-1, 0)^+$; $\zeta_i = 1 - \mathbb{I}(\lambda_i = 0) ~\forall i$
\vspace{-0.2cm}
\ENDWHILE
\vspace{-0.2cm}
\RETURN \hspace{-0.25cm} the final executed schedule by STOCHOS, $\mathcal{S}^* = \{\mathcal{S}^{STH}_1,...,\mathcal{S}^{STH}_{j-1} \}$ 

\end{algorithmic}
\end{algorithm}

\section{Experimental Results} \label{s:results} 
This section starts with the experimental setup, followed by our results and analyses. 
\subsection{Experimental Setup} \label{sub:case_studies} 
We run our optimization model for $100$ experiments representing different weather conditions and electricity prices. For each experiment, Algorithm 1 is implemented, and a maintenance schedule is obtained. To be able to conduct extensive numerical evaluations, we set $N_{\mathcal{S}} = 50$ scenarios and consider $N_{\mathcal{I}}=5$ WTs with a planning horizon of $20$ days. 

Data and forecasts used in this case study are described in detail in Section 3.1. True RLs (unknown to the planner ahead of time) are set at $\lambda_1 = 2.0$, $\lambda_2 = 6.8$, $\lambda_3 = 11.5$, $\lambda_4 = 16.2$, $\lambda_5 = 21.0$ days. Point predictions of those RLs (available to the planner) are set at $\hat{\lambda}_1 = 4.0$, $\hat{\lambda}_2 = 6.1$, $\hat{\lambda}_3 = 13.2$, $\hat{\lambda}_4 = 6.8$, $\hat{\lambda}_5 = 23.8$ days, reflecting different error magnitudes (low for WTs 2, 3, and 5, relatively high for WTs 1, and 4). Maintenance times are set at $\tau_1 = 11$, $\tau_2 = 5$, $\tau_3 = 6$, $\tau_4 = \tau_5 = 4$ hours. Maintenance times, as well as true and predicted RLs are set in light of the summary statistics for minor repairs reported by \cite{dinwoodie2015reference, carroll2016failure}. Those, together with the real-world data about the wind speed, wave height, and electricity price, are used as input to the probabilistic forecasting and scenario generation framework described in Section 3.2.

Table \ref{tab:pars} shows our selection for the remaining parameters: $K$ and $\Phi$ are chosen following \cite{yildirim2017integrated}, $\mbox{B}$, $\Psi$, and $\mbox{Q}$ are selected based on the recommendations of \cite{maples2013installation}, while $\mbox{C}_1$, $\mbox{C}_2$ are assigned large values relative to ``regular'' maintenance resource parameters, $\mbox{B}$ and $\Psi$. Medium-sized crew transport vessels (CTVs), which are common for minor to medium repairs in OSW turbines, are used, wherein $\Omega$, $\nu_{max}$, and $\eta_{max}$ are chosen according to the analyses in \cite{dalgic2013vessel,anderberg2015challenges}. Curtailment is assumed to be $0$\%, and hence, we set $C_{t,s} = 1~\forall t, s$. 
\begin{table}[]
   \caption{Operational parameters for our experimental setup}
    \renewcommand{\arraystretch}{.55}
    \centering
    \begin{tabular}{|c|c|c|}
      \hline
      Notation &    Parameter   &   Value   \\
      \hline
      K   &  Cost of preventive maintenance (PM) &   \$$4,000$\\
      $\Phi$   & Cost of corrective maintenance (CM)    &   \$$10,000$\\
      $\mbox{C}_1$, $\mbox{C}_2$    & Spot contracting costs &   \$$1,000$   \\
      $\mbox{B}$    &   Number of available crews &   $2$ crews   \\
      $\Psi$    &   Crew hourly rate    & \$$250$/hour  \\
      $\mbox{Q}$    &   Overtime hourly rate  &   \$$125$/hour \\
      $\mbox{W}$    &   Maximum number of non-overtime work hours   &   $8$ hours   \\
      H     &   Maximum number of overtime work hours & $8$ hours \\
      $\Omega$  &   Daily vessel rental cost  &    \$$2,500$/day  \\
      $\nu_{max}$   &   Wind speed safety threshold &   $15$ m/s    \\
      $\eta_{max}$  &   Wave height safety threshold &   $1.8$ m \\
      $t_R$     &   Time of first light & 6:00 am  \\
      $t_D$ &   Time of last sunlight   & 9:00 pm  \\
      \hline
    \end{tabular}
     \label{tab:pars}
\vspace{-0.2cm}
\end{table}

\subsection{Numerical Results} 
The performance of STOCHOS is compared against the following set of benchmarks: \vspace{-0.3cm}
\begin{enumerate}
    \item Holistic Opportunistic Strategy with Perfect Knowledge (\textbf{PK-HOST}): A deterministic variant of STOCHOS which assumes perfect knowledge of all parameters: metocean data, electricity prices, and RL predictions. This benchmark assesses how far STOCHOS is from the best possible performance attained had the planner known, with full certainty, all environmental and operational conditions. 
    \item Holistic Opportunistic Strategy with Point Forecasts (\textbf{PF-HOST}): A deterministic version of STOCHOS which uses the raw point forecasts of all the uncertain parameters: RU-WRF for wind speed, WAVEWATCH III for wave height, LEAR for electricity prices, and point RL predictions. This benchmark shows the value of uncertainty modeling in identifying the ``right'' windows of opportunity, and its relevance to fully harnessing the benefits of opportunistic maintenance grouping. 
    
    \item Condition-Based Strategy (\textbf{CBS}): A deterministic strategy that relies on point RL predictions, $\{\hat{\lambda}_i\}_{i \in \mathcal{I}}$, wherein maintenance is scheduled prior to the predicted RLs (with a three-day buffer time), or upon the occurrence of an unexpected failure. 
    \item Corrective Maintenance Strategy (\textbf{CMS}):  Maintenance tasks are only performed reactively, or post-failure. Both CBS and CMS are prevalent benchmarks in the academic literature and practice, and hence are used herein to showcase the combined benefit of accounting for opportunities and parameter uncertainties. 
\end{enumerate}

\begin{figure}[h]
    \centering
    \includegraphics[width=0.85\linewidth, trim= 0 1cm 0 0]{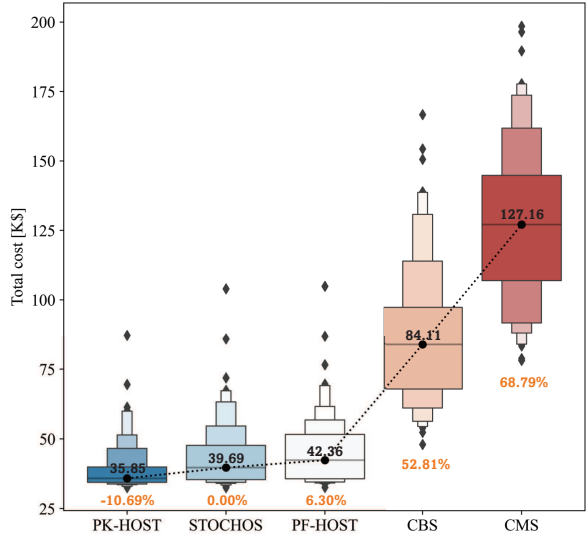}
    \caption{Boxplots of the total costs across $100$ optimization experiments corresponding to different metocean conditions and electricity price profiles. Solid circles (and the numbers on top of them) denote median costs (in \$K). Numbers below the boxplots are the percentage improvements (\%) of STOCHOS relative to each benchmark.} 
    \label{fig:cost_boxplots}
\end{figure}

Figure \ref{fig:cost_boxplots} illustrates the boxplots of the final costs (across $100$ experiments) achieved by STOCHOS and the four benchmarks, namely: PK-HOST, PF-HOST, CBS, and CMS. Looking at Figure \ref{fig:cost_boxplots}, we make few immediate observations: First, STOCHOS is the closest to PK-HOST with a relative cost difference (in terms of median costs) of $10.69$\%, while PF-HOST comes as a far second with a relative cost difference of $18.14$\%. This demonstrates how accounting for uncertainty makes the maintenance planner as close as possible to the ``utopian'' case of having perfect knowledge of information. 

Second, STOCHOS improves upon PF-HOST by $6.30$\%. Another interesting observation is that the interquantile range (the difference between the third and first quantiles) of STOCHOS is $24.16$\% smaller than that of PF-HOST, suggesting that STOCHOS is not only better ``on average,'' but is also \textit{consistently} better, and hence, a more reliable strategy. PF-HOST experiences multiple extreme cases where the performance is significantly worse than other strategies (e.g., the motivating example in Figure \ref{fig:intro_uncertainty}). This again confirms the value of accounting for uncertainty in terms of both average performance and reliability, relative to assuming full confidence in the available forecasts. Third, both STOCHOS and PF-HOST significantly outperform ``non-opportunistic'' approaches like CBS and CMS, further confirming the importance of opportunistic maintenance planning in OSW farms.

We summarize key O\&M metrics in Table \ref{tab:metrics}, including: number of vessel rentals, downtime, accessibility downtime (i.e. how much downtime is attributed to incorrect assessment of accessibility), production losses (in MWh) and revenue losses (in \$K). We also report the number of PM tasks (higher the better) and CM tasks (lower the better), as well as the number of maintenance interruptions. Finally, we report the total maintenance cost, as well as the increase in cost relative to the optimal solution. Across all metrics, STOCHOS significantly outperforms all benchmarks (excluding the non-realistic case of PK-HOST). 
\begin{table}[]
    \caption{O\&M metrics. Bold-faced values denote best performance (PK-HOST excluded)}
    \label{tab:metrics}
    \renewcommand{\arraystretch}{.62}
    \centering
    \begin{tabular}{l  c  c c c c}
    \hline
        & \textbf{PK-HOST} & \textbf{STOCHOS} & \textbf{PF-HOST} & {\textbf{CBS}} & \textbf{CMS} \\
        \hline
    {Number of vessel rentals} & $2.36$ & $\mathbf{2.18}$ & $2.26$ & {$4.92$} & $6.72$ \\
    {Total downtime [h]} & $39.52$ & \textbf{$\mathbf{44.20}$} & $47.30$ & {$97.20$} & $142.03$ \\
    {Accessibility downtime [h]} & $9.52$ & \textbf{$\mathbf{14.20}$} & $17.30$ & {$67.20$} & $112.03$ \\
    {Production loss [MWh]} & $101.21$ & \textbf{$\mathbf{141.05}$} & $154.59$ & {$597.68$} & $1037.24$ \\
    {Revenue loss [\$K]} & $5.10$ & \textbf{$\mathbf{7.10}$} & $7.83$ & {$29.85$} & $51.64$ \\
    {Total PM tasks} & $5.00$ & \textbf{$\mathbf{4.53}$} & $4.48$ & {$2.35$} & $0.00$ \\
    {Total CM tasks} & $0.00$ & \textbf{$\mathbf{0.47}$} & $0.52$ & {$2.65$} & $5.00$ \\
    {Maintenance interruptions} & $0.66$ & \textbf{$\mathbf{0.43}$} & $0.64$ & {$1.23$} & $1.83$ \\
    {Avg. total cost [\$K]} & $38.92$ & $\mathbf{43.43}$ & $45.00$ & {$86.90$} & $127.91$ \\
    {Cost inc. from optimal [\$K]} & $0.10$ & $\mathbf{4.60}$ & $6.14$ & {$48.07$} & $89.08$ \\
    {Median total cost [\$K]} & $35.86$ & $\mathbf{39.69}$ & $42.36$ & {$84.11$} & $127.16$ \\
    
\hline
    \end{tabular}
\end{table}

\subsection{The value of probabilistic forecasting} 
We would like to demonstrate the value of our probabilistic forecast and scenario generation approach presented in Section 3.2. To do so, we add two additional benchmarks: 
\begin{enumerate}
    \item Holistic Opportunistic Strategy with Calibrated Point Forecasts (\textbf{CPF-HOST}), which is similar to PF-HOST, except that it uses the predictive means of the probabilistic models as point predictions, instead of the raw point forecasts as in PF-HOST. 
    \item Stochastic Holistic Opportunistic Scheduler with Marginal Densities (\textbf{MD-STOCH OS}), which is similar to STOCHOS except that it uses scenarios generated from Gaussian distributions estimated via the residuals of the raw point forecasts of Section 3.1. This approach is prevalent in the classical stochastic optimization literature. 
\end{enumerate}

Table \ref{tab:disc} shows the median cost of those two benchmarks, in addition to that of STOCHOS and PF-HOST. The key interesting finding from this analysis is to show how an inadequate approach of handling uncertainty via marginal densities 
can lead to decisions that are even worse than deterministic strategies. Another interesting finding is that if the maintenance planner decides not to use the rich information in our probabilistic models (perhaps because of reluctance to change in business practice), then using the single-valued predictive means from our probabilistic models as input to the deterministic approach provides a modest but noticeable improvement in cost reduction. This is evident in how CPF-HOST improves upon PF-HOST. Finally, STOCHOS, with its adequate treatment of uncertainty and calibrated forecasts is able to achieve the maximal economic gain relative to both its deterministic and stochastic variants: PF-HOST, CPF-HOST, and MD-STOCHOS.  
\begin{table}[]
    \caption{The economic value of our probabilistic forecast framework.}
    \label{tab:disc}
    \renewcommand{\arraystretch}{.62}
    \centering
    \begin{tabular}{l C{2.4cm} C{2.4cm} C{2.4cm} C{3.2cm}}
    \hline
    & \textbf{STOCHOS} & \textbf{PF-HOST} & \textbf{CPF-HOST} & \textbf{MD-STOCHOS} \\
        \hline
    Median cost [\$K] & $\mathbf{39.69}$ & $42.36$ & $42.18$ & $44.57$ \\
    \hline
    \% Improvement [\%] & - & $6.30$ & $5.91$ & $10.95$ \\
    \hline
    \end{tabular}
\end{table}

\subsection{Sensitivity analysis and practical insights} 
\begin{figure}[]
    \centering
    \includegraphics[width=12cm, trim= 0 1cm 0 1cm]{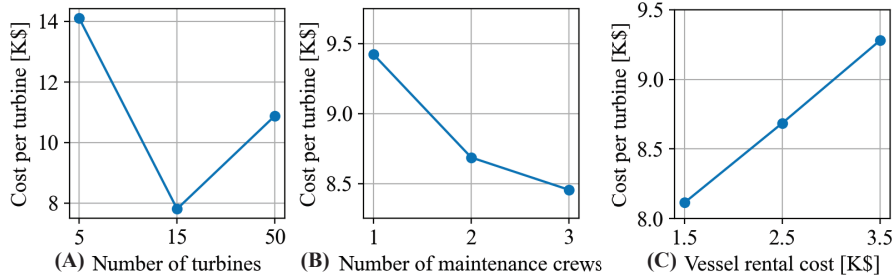}
    \caption{Costs resulting from the sensitivity analysis of the number of WTs (A), the number of available maintenance crews (B) and the transportation cost (C).}
    \label{fig:sens_analysis}
\end{figure}

To examine the effect of key operational parameters on the optimal solution, we conduct an experiment where we vary the number of WTs ($N_{\mathcal{I}}=5$, $15$ and $50$), maintenance crews (B $=1$, $2$ and $3$) and vessel costs ($\Omega= \$ 1,500$, $\$ 2,500$ and $\$ 3,500$). Figure \ref{fig:sens_analysis} shows the results, where we fix all experimental parameters and data as those of Section 5.1 (except the parameter being varied, be it $N_{\mathcal{I}}$, B, or $\Omega$). The only exception is that we use of B = $4$ crews (instead of B $= 2$) only when we vary the number of WTs (Figure \ref{fig:sens_analysis}A), since larger number of WTs (i.e. maintenance tasks) naturally requires higher maintenance resources.

Figure \ref{fig:sens_analysis}A shows a nonlinear relationship between the number of WTs and the maintenance cost per turbine. Analyzing the resulting schedules in more depth provides a possible explanation: with $5$ WTs and B $= 4$, the maintenance schedule is mainly dictated by the WT which is bound to fail first. In this case, WT1 with the shortest RL, is prioritized to avoid its failure. STOCHOS then finds it more economical to group the remaining tasks at the same day to share the setup cost already incurred by the maintenance of WT1, despite the production losses being relatively high. Going from $5$ WTs to $15$ WTs (with B $=4$) creates more room for cost savings, as not all maintenance tasks can be scheduled in the same day, which incentivizes the search for a better opportunity. 
With $N_{\mathcal{I}} = 50$ WTs, the schedule becomes too congested and some compromises must be made in order to complete all tasks on time and avoid the high costs of turbine failures.

Figure \ref{fig:sens_analysis}B shows that a too small  maintenance budget (e.g., only having access to B $= 1$ crew), results in higher maintenance costs, mainly due to prolonged maintenance schedules. Gradually increasing B parallelizes the maintenance workload and allows the sharing of other on-site maintenance resources (e.g., vessel rentals). This results in more efficient (i.e. shorter) maintenance schedules, although at the expense of higher crew costs. Finally, Figure \ref{fig:sens_analysis}C shows a fairly linear relationship between the total cost and the vessel rental costs. While the number of vessel rentals may be reduced in response to the increase in vessel costs, it causes an increase in the revenue losses and repair costs, since transportation-based opportunities are now weighted more heavily than their revenue-based counterparts.

Before we conclude, we would like to furnish few insights for OSW practitioners. First, one of our key findings in this work is the significance of accessibility information in OSW maintenance planning. Our research therefore calls upon both the research community and practitioners to develop more advanced, OSW-tailored access forecasting solutions \citep{gilbert2021probabilistic}, similar to those historically developed for classical maintenance parameters (e.g., asset degradation). Second, this work amplifies the recent calls in the academic community to shift away from deterministic to probabilistic forecasting and decision-theoretic frameworks \citep{pinson2013wind}. This is evident by the substantial improvements attained by STOCHOS relative to deterministic strategies like PF-HOST or CBS which solely rely on single-valued predictions in making critical OSW maintenance decisions.

\section{Conclusion}\label{s:conclusion} 
The unique challenges and uncertainties in OSW farms motivate the need for an offshore-tailored approach for maintenance optimization. To that end, we proposed STOCHOS, short for the stochastic holistic opportunistic scheduler\textemdash a maintenance scheduling approach that harnesses the maintenance opportunities arising due to favorable weather conditions, on-site maintenance resources, and maximal operating revenues, while adequately accounting for key operational and environmental uncertainties in the planning horizon. Our results showed that STOCHOS outperforms several prevalent benchmark strategies, across several key O\&M metrics. Future work will look into two broad research questions: (1) Given the same optimization model, can we seek more powerful probabilistic representations that can achieve maximal economic gains?; and (2) Given the same forecasting model, how can we improve our optimization models to accommodate more intricate representation of the complex failure modes in wind turbines, as well as the impact of vessel routing and crew logistics in large-scale wind farms. 

\section*{Supplementary Materials}
SM-1 describes how to statistically obtain wind power estimates given hub-height wind speed forecasts. SM-2 details the estimation of mission times given accessibility information. 

\section*{Data Availability Statement}
Details to reproduce the results of this work are included in the reproducibility report. The codes and data are available on GitHub: \url{https://github.com/petros-pap/STOCHOS}. 

\section*{Acknowledgements}
This work has been partially supported by the Rutgers Research Council Grants Program and the National Science Foundation (Grant \#: ECCS-2114422) 


\bibliographystyle{chicago}
\spacingset{.87}
\bibliography{IISE-Trans}
\end{document}